# Stimuli-responsive assembly of iron oxide nanoparticles into magnetic flexible filaments

**Aline Grein-Iankovski[1,2]\*, Alain Graillot[3], Watson Loh[2] and Jean-François Berret[1]\***

[1]*Université de Paris, CNRS, Matière et systèmes complexes, 75013 Paris, France.*
[2]*Institute of Chemistry, University of Campinas (UNICAMP), P.O. Box 6154 13083-970 Campinas, Brazil.*
[3] *Specific Polymers, ZAC Via Domitia, 150 Avenue des Cocardières, 34160, Castries, France.*

**Abstract**: The combination of multiple functionalities in a single material is an appealing strategy for the development of smart materials with unique features. In this work, we present the preparation of thermoresponsive magnetic nanoparticles and their one-dimensional assembly into transient microfilaments. The material is based on 9.4 nm iron oxide nanoparticles grafted with poly(*N-n*-propylacrylamide) *via* multiphosphonic acid anchoring sites. The hybrid nanoparticles present a low critical solution temperature (LCST) transition between 21 °C and 28 °C, depending on the pH and the ionic strength. When heated above the LCST in defined conditions, the nanoparticles aggregate and respond to an external magnetic field. An intrinsic characteristic of the thermoresponsive particles is an asymmetric transition between cooling and heating cycles, that was favorably exploited to build one-dimensional permanent microstructures, such as magnetic microfilaments and cilia. In summary, we present the development of a nanoplatform responsive to multiple stimuli, including temperature, magnetic field, pH and ionic strength and its transformation into magnetically active microfilaments that could find potential applications in remotely controlled devices.

**Keywords**:
Magnetic nanoparticles – Thermoresponsive Polymers - Self-assembly – Magnetic Filaments

## 1. Introduction

The development of smart materials has received great attention and effort, evolving into an important and continuously expanding field in areas such as nanomaterials, nanomedicine and biotechnology [1, 2]. Smart materials refer to compounds that respond to an external stimulus or to the environment with a dynamic and controlled change on their properties. There is a wide range of possibilities within the field of nanomaterials for the engineering of platforms sensitive to different stimuli, including pH, ionic strength, chemical composition, temperature, light, magnetic and electrical fields [3, 4]. Moreover, the combination of multiple functionalities, activated by different triggers, enables the tuning of the material properties in several ways, improving their performance and expanding the range of potential applications [5].

Temperature and magnetic field are compelling alternatives for external stimulus because they can be applied in a non-invasive manner [6]. When combined, e.g. in core-shell thermoresponsive superparamagnetic structures, the nanoparticles can be remotely guided with a magnet, depending on their aggregation state, which is in addition controlled by temperature [7–9]. In this case, the thermoresponsive polymer shell provides additional functionalities with respect to the colloidal stability and modulation of the magnetic core function. This is possible thanks to a strong temperature-dependent change in the hydration state of the polymer, which by contraction of the shell strengthens attractive interactions such as van der Waals and magnetic dipolar interactions, and favors the aggregation of the nanoparticles [10]. An additional level of control can be achieved with the presence of ionizable groups along the polymers, being then sensitive to the environmental pH and ionic strength of the dispersion. Both stimuli can play the role of internal triggers and influence the nanoparticle colloidal stability and aggregation state, via the control of the surface charge density [2].

Water soluble thermoresponsive polymers predominantly have a low critical solution temperature (LCST), which means they present a phase-transition from a soluble to an aggregate state upon heating to $T > T_{LCST}$ [11]. Poly(acrylamide) polymers are well-known examples of thermoresponsive materials. In this category, PNIPAm (poly(*N*-isopropylacrylamide) is certainly the most studied since the pioneering work of Heskins and Guillet [12, 13]. Whereas PNIPAm exhibits an LCST-type phase transition around 32 °C [12], its structural isomer poly(*N-n*-propylacrylamide) (PNnPAm) presents an LCST around 23 °C, which is an addition characterized by a steeper transition upon heating and a more pronounced heating/cooling hysteresis [14, 15]. These characteristics can be interesting for applications that benefit from a lower energy input to induce the phase transition or in memory devices that respond to temperature. Graillot *et al*. have successfully synthesized thermoresponsive copolymers of PNnPAm bearing phosphonic acid moieties (poly(*N-n*-propylacrylamide-*stat*-hydrolyzed(dimethoxyphosphoryl)methyl 2-methylacrylate), abbreviated PAmPh) that were assessed as polymeric sorbents of metal ions from aqueous effluents in low energy-consuming separation processes [16]. In our group, we have also demonstrated in previous reports the high affinity of phosphonic acid groups to metal oxide surfaces, including aluminum, iron, cerium and titanium oxides [22]. In such cases, the phosphonic acid moieties were found to adsorb spontaneously to the substrates through condensation of P-OH with hydroxyl groups of the metal surface and/or through the coordination of the phosphoryl oxygen to Lewis acid sites [17, 18]. Poly(ethylene glycol) copolymers containing multiple phosphonic acid have been shown to provide a resilient coating and promote long-term stability to the nanoparticles in biofluids and protein rich culture media [19–22].

Herein we applied the multiple phosphonic acid approach, developed for the coating of metal oxide nanoparticles, to coat iron oxide nanoparticles ($\gamma$-Fe$_2$O$_3$) with PAmPh, the objective being the design of novel magnetic nanoparticles with temperature regulated assembly properties. We further study the effect of heating-cooling hysteresis of the thermoresponsive polymer to build transient one-dimensional magnetic microfilaments. In a recent publication [23], we showed the potential of these hybrid nanoparticles to fabricate thermoresponsive magnetic cilia using a template-free approach. The prepared cilia, of lengths between 10 to 100 μm and aspect ratios up to 100, combined high magnetic susceptibility and mechanical flexibility, this latter having been associated to a persistence length of 10 μm [23]. Here, we describe the physicochemical characterization of the γ-Fe$_2$O$_3$@PAmPh nanoparticles in different conditions. We highlight the role of pH and ionic strength on the particle thermoresponsive behavior and demonstrate their ability to trigger the γ-Fe$_2$O$_3$@PAmPh assembly into flexible filaments.

## 2. Methods

### 2.1. Materials

The coating copolymer was custom synthesized by the company SPECIFIC POLYMERS (Montpellier, France) and consist of a statistical propyl acrylamide copolymer containing phosphonic acid groups, abbreviated PAmPh in the following (Fig. 1a). The synthesis was performed by free radical polymerization of thermosensitive *N-n*-propyl acrylamide monomers (NnPAm [25999-13-7] - SP-43-0-002) and (dimethoxy phosphoryl)methyl 2-methylacrylate (MAPC1, [86242-61-7] - SP-41-003) monomers in a molar ratio (98:02), using AIBN as a radical initiator [24]. The copolymer has a number-averaged molecular weight of $\overline{M}_n = 49500 \; g \; mol^{-1}$ with a dispersity of Đ = 2.8, *i.e.* approximately 8 phosphonic acid groups per polymer chain. The PAmPh thermal properties were monitored by turbidimetry. Fig. 1b displays the transmittance obtained from 0.1 wt.% dispersions at different pH values between 2.0 and 10.0 in the temperature range 18-35 °C. The continuous decrease of the transmittance observed in Fig.1b corresponds to the hydrophobically-driven aggregation of the polymer chains. The LCST critical temperatures were determined from the minimum of the first derivative, and these values were found to increase with pH, from 24 °C at pH 2.0 to 27 °C at pH 10.0. Sodium chloride (NaCl), nitric acid (HNO$_3$) and ammonium hydroxide (NH$_4$OH) were purchased from Sigma-Aldrich and purified water was obtained from a Milli-Q purification system.

The iron oxide nanoparticles were synthesized according to the Massart method [25] by alkaline co-precipitation of iron(II) and iron(III) chloride salts, which was followed by oxidation of the magnetite (Fe$_3$O$_4$) into maghemite (γ-Fe$_2$O$_3$) nanoparticles and size sorting. The resulting iron oxide dispersions were obtained at a weight concentration of 10 wt.% and at pH 1.8. The particles were positively charged, with nitrate as counterion. Under these conditions, the electrostatic interparticle repulsion ensured colloidal stability over a long period of time (> years). The size distribution of the uncoated γ-Fe$_2$O$_3$ particles was determined from transmission electron microscopy (Fig. 1c), leading to a median diameter of 9.4 nm and a dispersity (ratio between standard deviation and average diameter) of 0.17 (Fig. 1d). The magnetization curves of concentrated dispersions were obtained by vibrating sample magnetometry and adjusted using a saturation magnetization of 3.5 x 10$^5$ A m$^{-1}$ [20, 26].

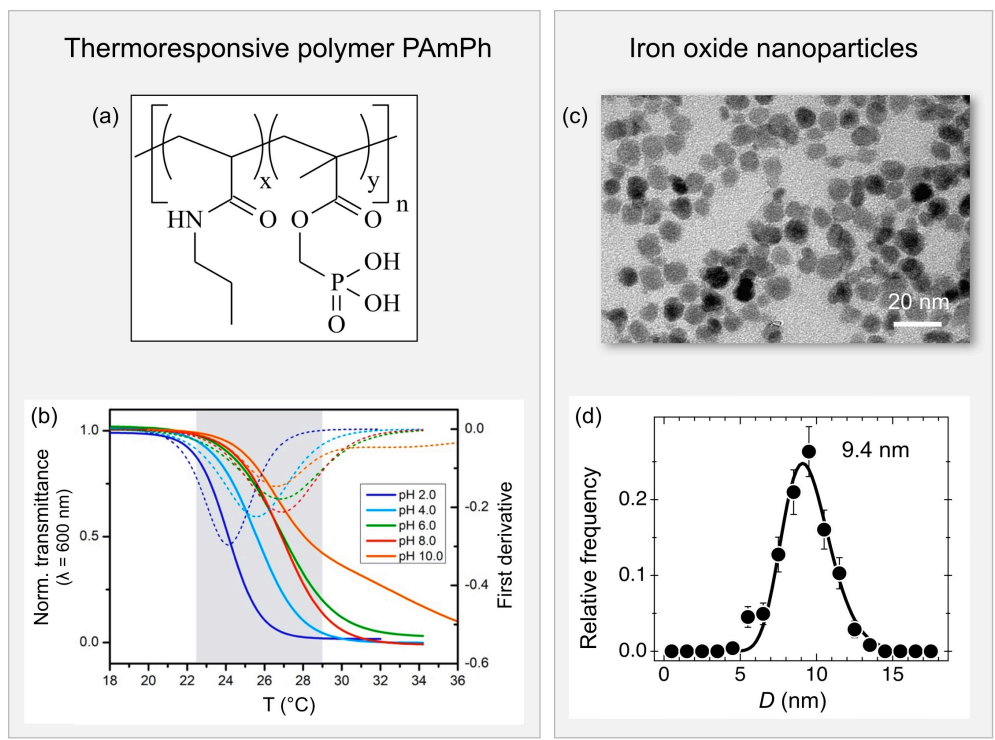

*Figure 1: a) Chemical structure of the PAmPh copolymer. b) Thermoresponsive behavior of the PAmPh copolymer as a function of pH: the solid lines show the normalized transmittance measured at λ = 600 nm and the dashed lines the respective first derivatives as a function of the temperature. c) Transmission electron microscopy (TEM) of the bare γ-$Fe_2O_3$ nanoparticles. d) Average size distribution of the bare γ-$Fe_2O_3$.*

## 2.2. Determination of the Nanoparticle to Polymer Coating Ratio

The first step for the preparation of thermoresponsive magnetic nanoparticles was the determination of the optimum amount of PAmPh needed to coat the γ-$Fe_2O_3$ surface in order to result in a stable colloidal dispersion. The procedure was based on a previous protocol developed in our group [20]. It consists in preparing PAmPh and γ-$Fe_2O_3$ dispersions at the same concentration (c = 0.1 wt.%), same pH (pH = 2.0), and in this case keeping the temperature below 10 °C. This technique avoids the nanoparticle aggregation upon mixing due to pH or salinity gap or due to temperature-induced phase transition, respectively. After overnight homogenization, the solutions were filtered with a 0.22 μm Millipore membrane. The γ-$Fe_2O_3$ dispersion was added dropwise to the PAmPh solution under magnetic stirring in an ice bath. The mixing was performed at different volume ratios $X = V_{NP}/V_{Pol}$, where $V_{NP}$ and $V_{Pol}$ are the respective volumes of nanoparticles and polymers. X was varied from 0.01 to 20.0, exploring a wide range of the phase diagram and keeping the total volume and concentration constant. After mixing, the pH was raised to 8.0 by the addition of $NH_4OH$ and the final dispersion stability was evaluated first visually and then by dynamic light scattering (DLS) measurements. The critical ratio $X_C$, here found at 0.5 was defined as the minimum amount of PAmPh to ensure γ-$Fe_2O_3$ colloidal stability at physiological pH [20, 21, 27].

### 2.3. Preparation of Thermoresponsive Magnetic Nanoparticles

After the determination of the critical $X_C$, a large batch (15 mL, 0.1 wt.%) of thermoresponsive magnetic nanoparticles was prepared, from which all subsequent experiments were performed. The formulation follows the same procedure as described in the last section. After raising the pH to 8.0 the prepared γ-$Fe_2O_3$@PAmPh nanoparticles were dialyzed with 100 kDa tubing membranes (Spectra/PorTM) for 4 days in the fridge at 4 °C to eliminate the excess of ions and non-linked polymers. Following this, the γ-$Fe_2O_3$@PAmPh dispersions were concentrated by centrifugation at 4000 rpm using Amicon filters unit of molecular weight cut-off 100 kDa (Merck) at 10 °C until a concentration of 0.1 wt% was reached. When necessary, the pH of the samples was adjusted with $HNO_3$ and $NH_4OH$ solutions.

### 2.4. Characterization Techniques

The hydrodynamic size and surface charge of the γ-$Fe_2O_3$@PAmPh were characterized by dynamic light scattering (DLS) and zeta potential analyses using a Malvern NanoZS Zetasizer instrument. The measurements were performed at 15 °C to avoid the temperature induced aggregation of the polymers and of the coated particles. The DLS technique was also used to characterize the thermal behavior of the γ-$Fe_2O_3$@PAmPh and its phase transition at different conditions. For this, 200 μL of each sample (0.1 wt% at pH 2.0, 4.0, 6.0, 8.0 and 10.0), first conditioned at 5 °C, was introduced in the NanoZS measuring chamber previously stabilized at 40 °C. The scattering intensity and hydrodynamic diameter were then recorded over time. The equipment was set to perform 90 measurements of 10 seconds each, resulting in a total analysis time of 15 minutes.

The determination of the salt effect on γ-$Fe_2O_3$@PAmPh thermal phase transition at different pH was also performed using DLS. 200 μL of each studied sample (0.1 wt% at pH 2.0, 4.0, 6.0, 8.0 and 10.0) was thermalized at 40 °C in the measurement chamber for 5 minutes. The scattering intensity and hydrodynamic diameter was monitored after successive additions of 2.0 mol $L^{-1}$ NaCl aliquots and a standing time of 3 minutes. The salt concentration was increased until the nanoparticles aggregate in large clusters and began to settle.

The determination of the LCST was performed by turbidimetry. The transmittance of each sample at a concentration of 0.1 wt.% was measured in a Jasco V-630 UV-Visible spectrometer equipped with a Peltier temperature controller. The temperature ramp was performed between 18 °C to 35 °C with a heating rate of 0.1 °C $min^{-1}$. The wavelength of 600 nm was selected to follow the transmittance transition. The LCST, considered as an approximation from the cloud point temperature, was determined from the minimum of the derivative curve.

### 2.4. Preparation of Magnetic Filaments

The magnetic filaments were prepared by the controlled assembly of the thermoresponsive magnetic nanoparticles under the influence of a magnetic field and triggered by temperature. 100 μL of γ-$Fe_2O_3$@PAmPh 0.1 wt% at pH 8.0 containing increasing amounts of NaCl (0, 10, 50 and 200 mM) were filtered (0.45 μm Millipore membrane), filled in a vial (1 mm thick, 1 cm wide) and placed between two Neodymium magnets (1 cm x 5 cm x 0.5 cm). The system was heated to 40 °C for 30 minutes, cooled down and analyzed using Optical Microscopy. The formed filaments were washed twice with Milli-Q water by magnetic decantation and finally redispersed in water.

**2.5. Optical Microscopy**

The filaments were analyzed using an IX73 Olympus inverted microscope equipped with 40x objective lens and coupled with a CCD camera (QImaging, EXi Blue). The images were obtained with the Metaview software (Universal Imaging) in the phase contrast mode and analyzed using the ImageJ software [28]. To visualize the magnetic response of the magnetic filaments and their motion, a rotating magnetic field of 10 mT was applied thanks to two pairs of coils working with a 90°-phase shift [29].

## 3. Results and Discussion

**3.1 Coating iron oxide nanoparticles with thermo-responsive polymers**

In this section, we describe the coating of 9.4 nm maghemite particles ($\gamma$-Fe$_2$O$_3$) by a statistical propyl(acrylamide) copolymer (PAmPh) containing phosphonic acid groups [20, 27]. Dispersions of nanoparticles and polymer solutions prepared at the same concentration ($c$ = 0.1 wt.%) and same pH (pH 2) were mixed at different volume ratios $X$. The mixing was executed at temperature T = 5 °C, *i.e.* below the LCST of the polymers. The pH of the dispersion was then increased to pH 8.0, while maintaining the temperature at 5 ° C. Fig. 2a displays images of the vials containing the dispersions for $X$ = 0.01 to 20. The color change of the dispersions is due to the changes in $\gamma$-Fe$_2$O$_3$ concentration, this concentration varying as $c_{NP} = cX/(1 + X)$. Likewise, along the $X$-line the polymer concentration decreases with $X$ as $c_{Pol} = c/(1 + X)$. Interestingly, it is found that above the critical ratio $X_C$, here for $X$ > 0.5 the mixed dispersions are turbid, and the particles aggregated. This result was further confirmed with dynamic light scattering measurements. Fig. 2b presents the hydrodynamic size distributions with increasing $X$. Below $X_C$, the NPs are characterized by a monomodal size distribution centered around 100 nm, whereas, above this value, the distributions are multimodal and shifted towards larger sizes (as e.g. for $X$ = 1.5). This outcome is interpreted by assuming that above $X_C$, the particles are insufficiently coated and cannot withstand the pH change to pH 8, as observed for bare particles [20, 30]. We exploit this feature to calculate the number of adsorbed chains per particle $n_{ads}$ given by the relation $n_{ads} = M_n^{NP}/(X_C M_n^{Pol})$ [20, 26], where $M_n^{NP}$ and $M_n^{Pol}$ are the number-averaged molecular weights of the particle and polymer, respectively. Assuming here $X_C$ = 0.5 (Fig. 2a), it is found that the particles are coated in average with 54 copolymer chains, which represents approximately 432 phosphonic acid groups per particle.

The coated iron oxide nanoparticles prepared at $X_C$ = 0.5 and at temperature below the LCST of the polymer ($T_{LCST}^{Pol}$ = 25 °C [23]) will be designated as $\gamma$-Fe$_2$O$_3$@PAmPh in the sequel of the paper. A schematic representation of the coated particles is show in Fig. 2c. Pertaining to colloidal stability, $\gamma$-Fe$_2$O$_3$@PAmPh have been found to be stable over time (> months) and over a broad range of pH values. DLS measurements performed on bare and coated $\gamma$-Fe$_2$O$_3$ (pH 2) reveal, for instance, hydrodynamic diameters $D_H$ = 25 nm ($pdi$ = 0.15) and $D_H$ = 91 nm ($pdi$ = 0.18) respectively (Fig. 2d). As the scattering intensity varies with the 6$^{th}$-power of the particle size (for a fixed particle number density), large particles contribute more strongly than the small ones, leading to the shift of the hydrodynamic size to larger values. In the present case, we surmise that the size increase from 25 to 91 nm seen in DLS is due to the combination of a partial aggregation of the initial particles and of broader size distribution. By varying the pH up to 10.0 the γ-Fe$_2$O$_3$@PAmPh remained also stable, exhibiting a slight increase in size. This size variation may be explained by the deprotonation of the phosphonic acid groups located in the corona when the pH exceeds the pKa values of the phosphonic acid (pKa$_1$ = 2.13, pKa$_2$ = 7.06) [31]. As shown in Fig. 2e, with increasing pH, γ-Fe$_2$O$_3$@PAmPh become more negative, with zeta potential increasing from -5 to around -30 mV. This behavior suggests the presence of unbound phosphonic acids in their corona. The PAmPh is a statistical copolymer, meaning that the 8 phosphonic acid groups per copolymer are randomly distributed along the chain. Due to steric and segment rigidity hindrance our results indicate that not all anchoring groups reach the metallic surface and contribute to the polymer binding. The long-term stability shown by γ-Fe$_2$O$_3$@PAmPh (more than one year) suggests however that the grafting is achieved *via* multiple site binding [22].

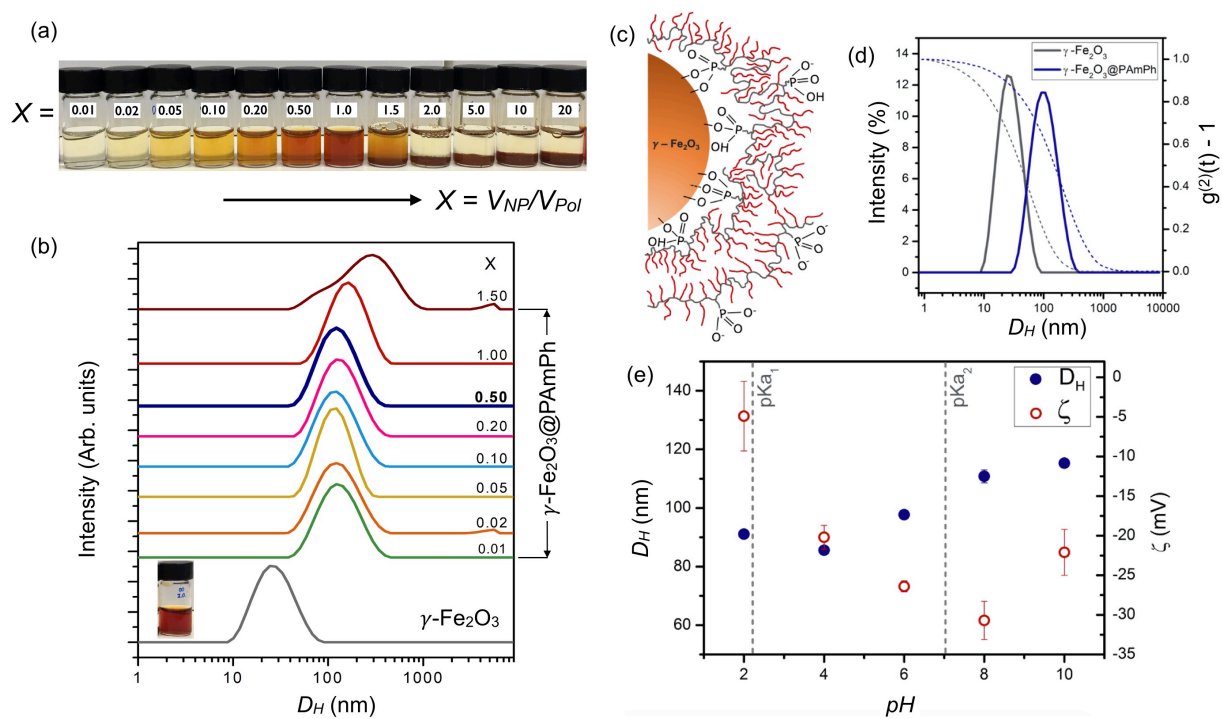

***Figure 2:*** *a) Images of vials containing γ-Fe$_2$O$_3$ and PAmPh polymers at different X, where X is defined as the ratio between the volumes of nanoparticle and polymer dispersions (pH 8.0). Above a critical ratio X$_C$, here comprised between 0.5 and 1, the dispersions are turbid and the particles agglomerate.* ***b)*** *Hydrodynamic size distributions for the dispersions shown in Fig. 2a obtained at 15 °C. From these data, the critical mixing ratio X$_C$ = 0.5 was derived (thick blue line). X = 0.5 also corresponds to the condition chosen for the preparation of the γ-Fe$_2$O$_3$@PAmPh batch used in the sequel of the paper.* ***c)*** *Schematic representation of the structure of the polymer coating layer around the magnetic nanoparticles.* ***d)*** *Light scattering autocorrelation function $g^{(2)}(t) - 1$ (dashed lines) and respective hydrodynamic diameter distributions (continuous line) of bare γ-Fe$_2$O$_3$ and γ-Fe$_2$O$_3$@PAmPh at pH 2.0.* ***e)*** *Hydrodynamic diameter D$_H$ and zeta potential variation of γ-Fe$_2$O$_3$@PAmPh particles as a function of the pH. The dashed lines indicate the pKa$_1$ and pKa$_2$ values of phosphonic acid group (pKa$_1$ = 2.7 and pKa$_2$ = 7.8) [20].*

### 3.2 Electrostatics prevents temperature-induced nanoparticle assembly

Earlier studies have shown that the presence of hydrophilic monomers or charged groups in thermoresponsive polymer chains increases its solubility in water-borne solvents, resulting in a higher LCST [10, 32]. Likewise, here, the observed LCST of pure PAmPh solutions was found to gradually shift to higher temperatures, from 24 °C to 27 °C, as the ionization degree of the phosphonic acid groups increased with pH variation, from 2.0 to 10.0 (Fig. 1b). This outcome has a direct impact on γ-$Fe_2O_3$@PAmPh thermo-responsiveness towards temperature changes. In a second series of experiments, the γ-$Fe_2O_3$@PAmPh dispersions were subjected to a steep temperature jump, from 5 to 40 °C. It is anticipated that above the LCST, the particles should aggregate as the result of preferential polymer-polymer interactions and corona shrinkage. Fig. 3 shows values of the observed time dependent hydrodynamic diameters ($D_H$) and scattered intensity ($I_S$) obtained from DLS measurements. These experiments were performed at 5 different pH values between 2 and 10. At pH 2.0 a rapid increase in the $D_H$ is observed, indicating a transition from a dispersed to an aggregated state, and followed by the rapid sedimentation of the aggregates. This sedimentation is revealed by the decrease of the intensity observed in Fig. 3b two minutes after the T-jump. Above pH 4.0 however, *i.e.* when the nanoparticles are significantly charged, $D_H$ values remained constant over the entire time range (Fig. 3a).

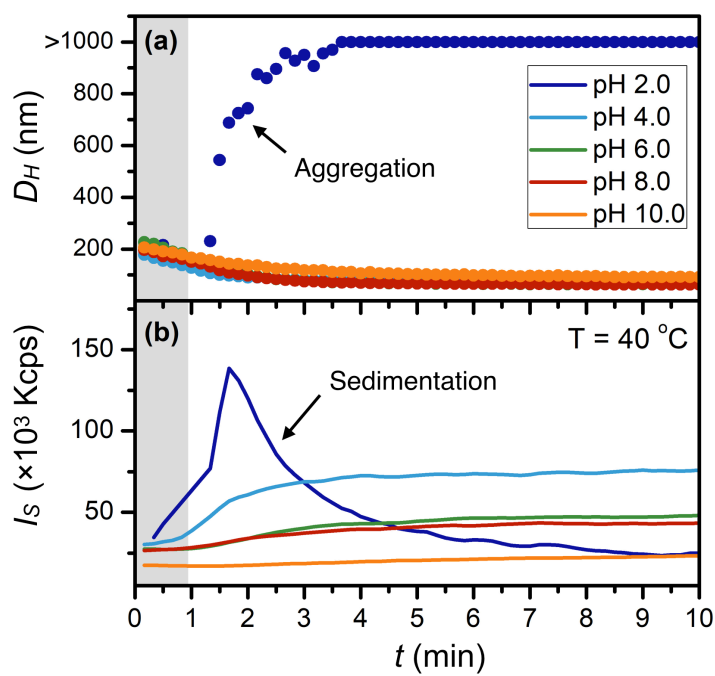

*Figure 3:* Time dependence of the (a) hydrodynamic diameter $D_H$ and (b) light scattering intensity $I_S$ obtained from γ-$Fe_2O_3$@PAmPh dispersion after a temperature jump from 5 to 40 °C. The initial grey region accounts for the initial temperature stabilization period.

Overall, no significant aggregation of the γ-$Fe_2O_3$@PAmPh at pH 4.0, 6.0, 8.0 and 10.0 was observed even up to 70 °C (data not shown). Note that the initial decay in $D_H$ over the first minute (grey area) is due to the sample thermalization, and to the fact that the solvent viscosity was fixed to that of the solvent at 40 °C throughout the experiment. These results suggest that the amount and distribution of phosphonic acid groups affect the copolymer thermal response, in agreement with earlier reports [24, 31]. In particular, it was found that the changes in LCST are more pronounced in statistical copolymers compared to block copolymers. PAmPh statistical copolymers also lose their thermoresponsive behavior when the phosphonic acid moieties are increased to around 30% of the chain monomers [24]. An additional and more challenging factor is imposed when such polymers are grafted at the curved surface of a nanoparticle [33]. One or more of the effects listed above could account for the non thermo-responsiveness of the $\gamma$-$Fe_2O_3$@PAmPh particles at elevated pH [10].

### 3.3 Triggering thermo-responsiveness using salt

To overcome this limitation, we investigated the possibility to trigger the $\gamma$-$Fe_2O_3$@PAmPh thermoresponsivity by modulating the interparticle electrostatic potential through addition of salt. With this aim, dispersions of coated nanoparticles were prepared at increasing NaCl concentrations, between 10 and 1000 mM and at pH 4.0, 6.0, 8.0 and 10.0. Fig. 4 shows DLS hydrodynamic diameters (Fig. 4a) and scattering intensity (Fig. 4b) of $\gamma$-$Fe_2O_3$@PAmPh nanoparticles as a function of the NaCl concentration. Each data point was measured three minutes after stabilization at 40 °C. At low salt concentration, both $D_H$ and $I_S$ are found to increase, the increases being shifted to higher salt concentration when pH changes from 4.0 to 10.0. Moreover, the scattering intensities displays pH-dependent maxima (Fig. 4b). These maxima are due to the sedimentation of the large aggregates formed, and to the subsequent reduction of particle concentration in the scattering volume. This peak is interpreted as the critical NaCl concentration required to reduce the electrostatic repulsion between $\gamma$-$Fe_2O_3$@PAmPh nanoparticles and to enable an effective temperature-induced assembly. These NaCl concentrations are 50, 140, 200 and > 600 mM for pH 4.0, 6.0, 8.0 and 10.0, respectively. The magnitude of this behavior is also dependent on the surface charge of the nanoparticle, being more evident at pH 4.0 and gradually mitigated at higher pH.

### 3.4 Coated nanoparticle thermoresponsive behavior

Based on the previous results, we now investigate the phase transition as a function of temperature for the following conditions: pH 2.0 (no added salt), pH 4.0 with 50 mM NaCl, pH 6.0 with 140 mM NaCl and pH 8.0 with 200 mM NaCl. The light transmittance measured at $\lambda = 600$ nm was recorded during heating between 18 to 35 °C (Fig. 5). The values of the LCSTs were determined from the derivative of the transmittance with respect to the temperature (dotted curves in Fig. 5). Here we assume that the minimum found in the derivatives is indicative of the sample LCST. We observe that the LCST of $\gamma$-$Fe_2O_3$@PAmPh is shifted to higher temperatures as the ionization degree of the polymer shell increases, varying from 21.5 °C at pH 2.0 to 27.3 °C at pH 8.0 with 200 mM of NaCl. When compared to the LCST of pure PAmPh solution at pH 2.0 (Fig. 1b), we found a moderate decrease of the LCST from 24 °C to 21.5 °C. This effect was already observed in other thermoresponsive nanoparticles, and was ascribed to the solvation inhibition of the chain end anchored at the nanoparticle surface [10]. Above pH 4.0, the shift to lower temperatures was not observed, possibly compensated by the opposed effect from the presence of charged groups.

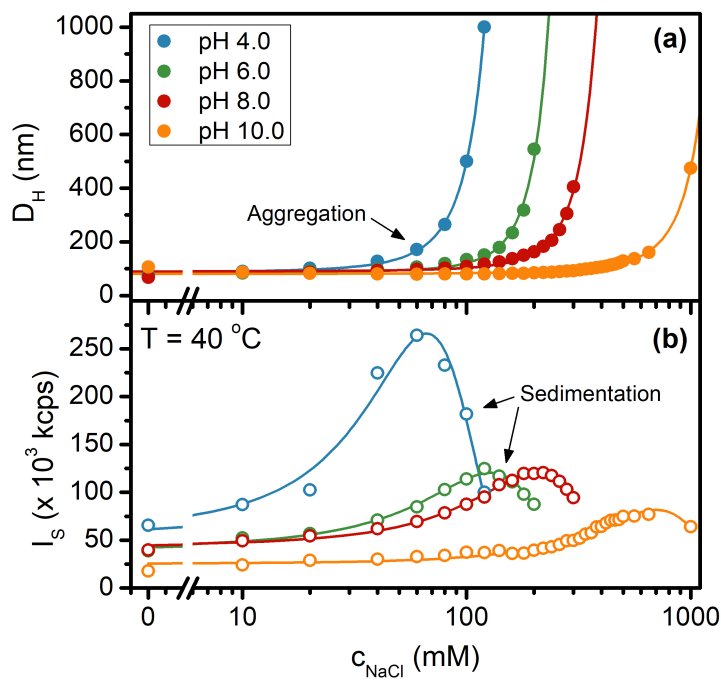

***Figure 4:*** *Thermoresponsive assembly of the γ-Fe$_2$O$_3$@PAmPh at different pH as a function of NaCl concentration. (a) Hydrodynamic diameter $D_H$ and (b) light scattering intensity $I_S$. In this experiment, the samples were thermalized at 40 °C the measurements were performed after the addition of a NaCl aliquot and a standing time of 3 minutes for temperature stabilization.*

It is interesting to note that the transmittance decay seen in Fig. 5 becomes broader as the ionization degree of the coated particles increases. This broadening can be correlated with the variation in the local environment experienced by different monomers in the shell. For highly curved interfaces, two or more phase transitions have been identified using differential scanning calorimetry (DSC). This was found for gold nanoparticles grafted with PNIPAm [34] or iron oxide nanoparticles grafted with poly(2-isopropyl-2-oxazoline) [35]. The higher temperature peaks observed by DSC are generally attributed to the outer shell layer, consisting of highly hydrated and less densely packed chains [34, 35]. This assignment corroborates our results, indicating that the organization of the polymer chains exposing phosphonic acid groups to the outer layer significantly shifts the LCST to higher temperatures, broadening the transition range as a result of the ionization degree of the chain. The onset of the transition at the lowest temperature inflection point hence corresponds to the inner part of the polymer shell and shows a narrower variation with pH [35].

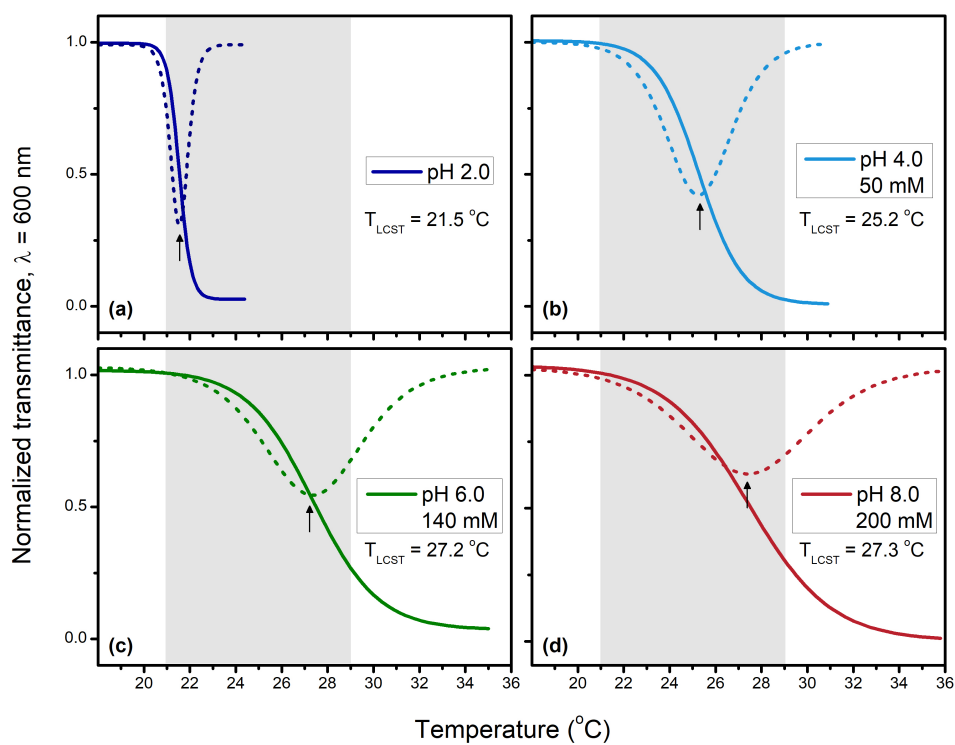

*Figure 5.* Determination of the low critical solution temperature (LCST) for γ-Fe$_2$O$_3$@PAmPh nanoparticles. Experimental conditions are a concentration of 0.1 wt.% and (a) pH 2.0, (b) pH 4 with 50 mM NaCl, (c) pH 6.0 with 140 mM NaCl, (d) pH 8.0 with 200 mM NaCl. The plots show the normalized transmittance measured at λ = 600 nm (continuous line) and its first derivative (dashed line) as a function of the temperature.

### 3.5 Heating and Cooling hysteresis

Another important aspect to consider for potential applications of thermoresponsive materials is the reversibility of their phase transition. In a subsequent experiment, we have studied the reversibility of the transition by measuring the hydrodynamic size $D_H$ and scattering intensity $I_S$ during a heating-cooling cycle. Fig. 6a shows the time dependence of the scattering intensity upon heating after a steep temperature jump from 5 to 40 °C. The increase in $I_S$ indicates that the γ-Fe$_2$O$_3$@PAmPh nanoparticles self-assembled under the effect of temperature. The shape of the curves thereby reveals a characteristic growth rate that is pH dependent and minimum for highly charged nanoparticles (Fig. 6a). Upon cooling to 10 °C, it is found that the particle $D_H$ decreases over time, indicating that the initially formed aggregates disassemble into single particles (Fig. 6b). The disassembly process occurs however in a time scale of several days, that is, much longer than that of the assembly. The results of Fig. 6 also show that the disassembling rate depends on the pH conditions. Surface charges play here an important role, contributing to a faster disassembly process observed at higher pH values. Yet, even after two weeks, γ-Fe$_2$O$_3$@PAmPh remains in a partially aggregated state, exhibiting a larger size than that found prior to heating. Note, however, that for nanoparticles at pH 10, the return to the initial state is complete after about two weeks at low temperature, whereas this process takes about one month at lower pH.

Some differences between the heating and cooling cycles have been reported in the literature, for both polymer solutions [36, 37] or hybrid thermoresponsive materials [35, 38, 39]. In these cases it was found that the transition enthalpy on heating is higher than during cooling, indicating a slower redispersion process [35, 36]. The main reason is generally attributed to the slower diffusion of water into the aggregated polymer network due to steric hindrance imposed by the entangled chains [10]. In addition, some parts of the polymer chain can undergo a conformation change during prolonged heating inducing non-reversible aggregation or crystallization [40]. In the case of PNnPAm, studies have shown that it presents a sharp temperature transition combined with a large hysteresis when compared to its structural isomer PNIPAm [14, 15, 41]. Similarly, molecular dynamics simulations have evidenced that PNnPAm has a stronger temperature dependence and a higher water coordination number around its hydrophobic side chain. This arrangement contributes to enhanced entropic effects associated with the release of the water molecules at lower temperatures, and results in a stronger interchain hydrophobic interaction [42].

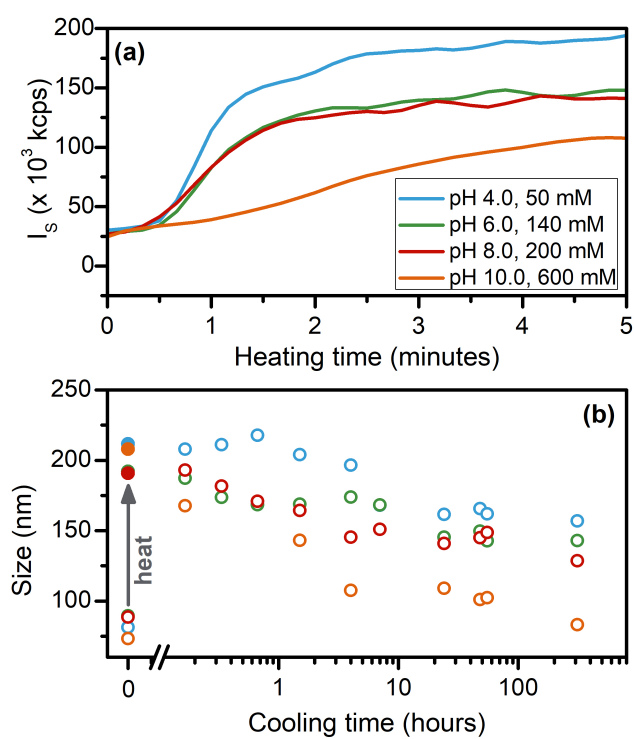

**Figure 6.** Assembly and disassembly of the $\gamma$-$Fe_2O_3$@PAmPh particles induced by the temperature at different pH and salt concentrations. (a) Heating: variation of the light scattering intensity ($I_S$) as a function of the time after a temperature jump from 5 to 40 °C. (b) Cooling: Hydrodynamic diameter ($D_H$) variation as a function of time after cooling down the system to 10 °C.

### 3.6 Nanoparticle assembly into flexible filaments

The slow redispersibility of thermoresponsive materials is generally regarded as a drawback for applications [43]. On the other hand, here, we show that the large hysteresis observed in Fig. 6 can represent a definite advantage for the design of nano- or micro-functional objects. In particular, it is conceivable that self-assembled microstructures can be formed in a few minutes and be persistent for days or weeks depending on the preparation conditions. One example of such advanced microstructures is shown in Fig. 7. These figures show optical microscopy images of magnetic filaments obtained by the concomitant application of temperature ($T >$ LCST) and of a constant magnetic field of 300 mT to a $\gamma$-Fe$_2$O$_3$@PAmPh dispersion. When the nanoparticles are exposed to an external magnetic field, strong and directional dipole-dipole interactions induce the formation of nanoparticle chains [44]. The concomitant heating promotes the phase-transition within the polymer shells, and favor the interparticle attraction and sticking. Interestingly, these filaments retain their configuration after the magnetic field is removed or the dispersion is cooled down to room temperature (Fig. 7b). The resulting magnetic filaments can be washed, isolated and used later for further experiments. As shown in Fig. 7c, the filaments keep their structures over time, and have inherited a certain flexibility from the polymers. They also exhibit a response to an applied magnetic field, and can rotate, bend, or move under the application of an external magnetic field, as can be seen in the Movie 1 (Supporting information).

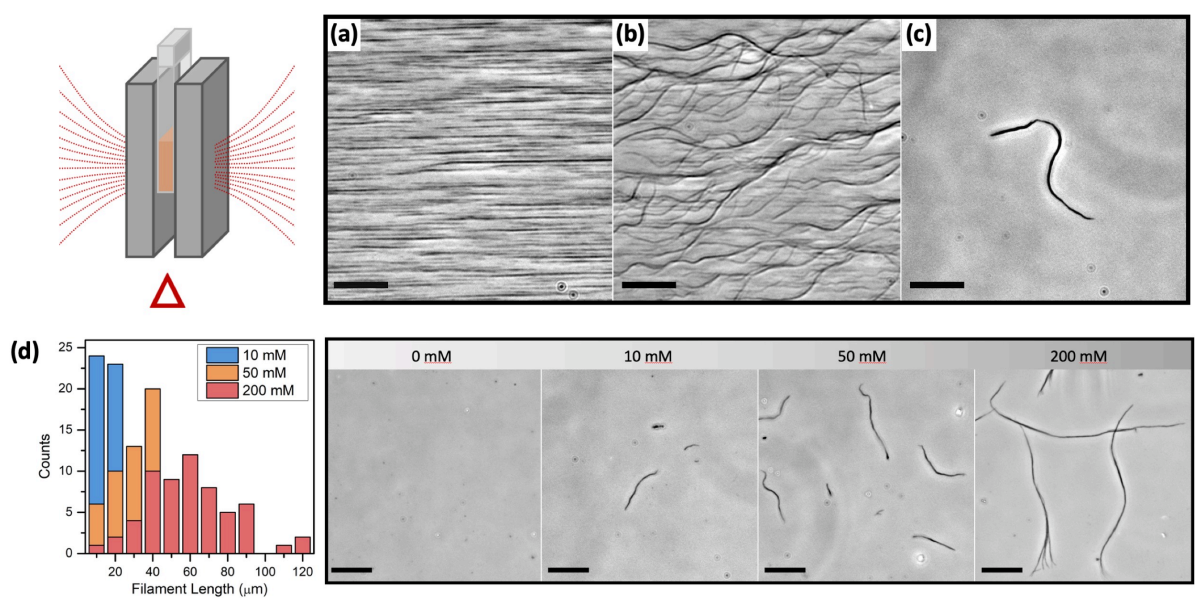

*Figure 7.* *Fabrication of magnetic filaments through the concomitant action of temperature and magnetic field using $\gamma$-Fe$_2$O$_3$@PamPh dispersions. The sample is placed between two magnets and the assembly is triggered by heating. Optical microscopy images of the (a) formed filaments under the application of the magnetic field, (b) magnetic filaments after the removal of the field, and (c) an isolated magnetic filament at room temperature after washing and separation. (d) Effect of the NaCl concentration on filaments formation at pH 8 and their length distribution. Scale bars are 20 μm.*

We demonstrated previously that the combination of physicochemical stimuli can act as internal triggers for the thermoresponsive behavior of the $\gamma$-Fe$_2$O$_3$@PAmPh (Fig. 4). These parameters can also be used to trigger the formation of the magnetic filaments and adjust their final dimensions. Fig. 7d shows the effect of salt on the formation of filaments from a $\gamma$-Fe$_2$O$_3$@PAmPh dispersion at concentration 0.1 wt. % and pH 8.0. In agreement with the results above, we observe that salt can control the one-dimensional growth of the filaments in a very gradual way. At the highest

NaCl content, it is found that the filaments are more numerous, longer and thicker. The average length derived from the distributions in Fig. 7d are 20 ± 11, 34 ± 14 and 60 ± 23 μm for filaments obtained at 10, 50 and 200 mM, respectively. The average filament diameter obtained at [NaCl] = 50 mM from scanning electron microscopy was found at 370 ± 96 nm. These flexible filaments are promising magneto-actuators that could find potential applications as microrobots, microstirrers, micromanipulators or microrheological probes [45]. In conjunction with the fabrication of magnetic cilia that we have demonstrated in a recent publication [23], these results shows the remarkable potential of the $\gamma\text{-}Fe_2O_3$@PAmPh as a multi-responsive nanoplatform for the fabrication of microstructured devices.

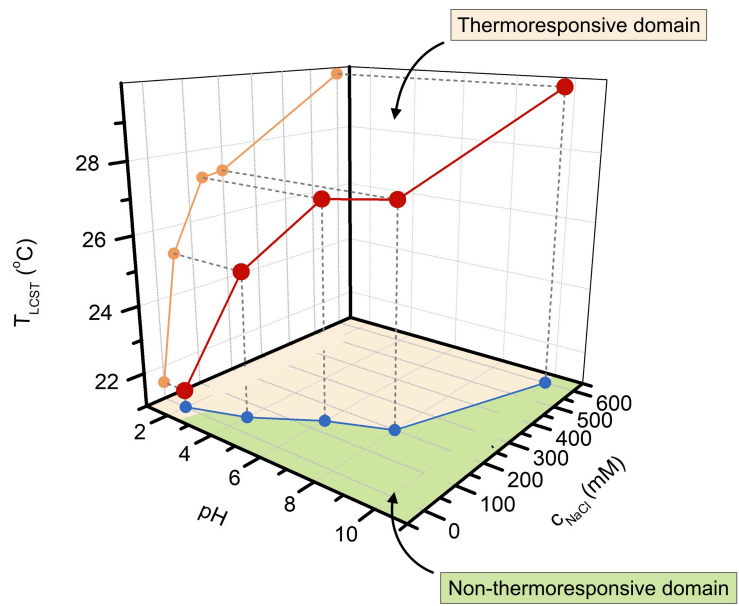

***Figure 8.*** *Diagram of the thermoresponsive behavior of $\gamma\text{-}Fe_2O_3$@PamPh summarizing the intrinsic correlation between pH, salt concentration and the LCST of the hybrid nanoparticles.*

## 4. Conclusion

In this work we report the synthesis and physicochemical characterization of thermoresponsive magnetic nanoparticles based on the grafting of PAmPh onto 9.4 nm iron oxide nanoparticles aiming at the development of a multi-responsive material. The resulting hybrid nanoparticles $\gamma\text{-}Fe_2O_3$@PamPh are shown to respond to internal as well as external triggers, such as pH, ionic strength, temperature and magnetic field. The polymeric shell around the particles dictates the thermoresponsive behavior of the hybrid structures. In particular, pH and ionic strength are parameters that control the $\gamma\text{-}Fe_2O_3$@PamPh aggregation state, in combination with a temperature jump through the LCST, as summarized in Figure 8. The presence of ionizable non-bound phosphonic acid groups in the polymeric shell makes the particles responsive to pH, changing their LCST behavior and hindering their thermoresponsive behavior. In this case the addition of salt can be used to modulate the interparticle electrostatic repulsion and to trigger the temperature induced phase transition. Screening the electrostatic interaction via salt addition leads to the nanoparticle aggregation above the LCST, increasing their responsiveness towards a magnetic field. The $\gamma\text{-}Fe_2O_3$@PamPh present a significative hysteresis during heating/cooling cycles that was favorably exploited to build magnetically-active flexible filaments that can find promising applications in remotely-controlled devices. It is therefore a system of great interest that provides different levels of control through the manipulation of pH, salt concentration, temperature and magnetic field.

## Acknowledgments

The authors thank the financial support provided by the São Paulo Research Foundation (FAPESP, grant numbers 2015/25406-5, 2017/04571-3, and 2018/16330-3). ANR (Agence Nationale de la Recherche) and CGI (Commissariat à l'Investissement d'Avenir) are gratefully acknowledged for their financial support of this work through Labex SEAM (Science and Engineering for Advanced Materials and devices) ANR 11 LABX 086, ANR 11 IDEX 05 02. We acknowledge the ImagoSeine facility (Jacques Monod Institute, Paris, France), and the France BioImaging infrastructure supported by the French National Research Agency (ANR-10-INSB-04, « Investments for the future »). This research was supported in part by the Agence Nationale de la Recherche under the contract ANR-13-BS08-0015 (PANORAMA), ANR-12-CHEX-0011 (PULMONANO), ANR-15-CE18-0024-01 (ICONS), ANR-17-CE09-0017 (AlveolusMimics) and by Solvay.

## Supporting Information

Movie 1: Different motion responses of magnetic filaments prepared from a 0.1 wt.% $\gamma$-$Fe_2O_3$@PAmPh dispersion under the stimulus of a rotating magnetic field.

## References

1. Stuart, M.A.C., Huck, W.T.S., Genzer, J., Müller, M., Ober, C., Stamm, M., Sukhorukov, G.B., Szleifer, I., Tsukruk, V. V, Urban, M., Winnik, F., Zauscher, S., Luzinov, I., Minko, S.: Emerging applications of stimuli-responsive polymer materials. Nature Materials. 9, 101–113 (2010). https://doi.org/10.1038/nmat2614

2. Lu, Y., Aimetti, A.A., Langer, R., Gu, Z.: Bioresponsive materials. Nature Reviews Materials. 2, 16075 (2016). https://doi.org/10.1038/natrevmats.2016.75

3. Jochum, F.D., Theato, P.: Temperature and light-responsive smart polymer materials. Chemical Society Reviews. 42, 7468–7483 (2013). https://doi.org/10.1039/C2CS35191A

4. Xia, F., Jiang, L.: Bio-Inspired, Smart, Multiscale Interfacial Materials. Advanced Materials. 20, 2842–2858 (2008). https://doi.org/https://doi.org/10.1002/adma.200800836

5. Guragain, S., Bastakoti, B.P., Malgras, V., Nakashima, K., Yamauchi, Y.: Multi-Stimuli-Responsive Polymeric Materials. Chemistry – A European Journal. 21, 13164–13174 (2015). https://doi.org/https://doi.org/10.1002/chem.201501101

6. Schmidt, A.M.: Thermoresponsive magnetic colloids. Colloid and Polymer Science. 285, 953–966 (2007). https://doi.org/10.1007/s00396-007-1667-z

7. Sun, Y., Ding, X., Zheng, Z., Cheng, X., Hu, X., Peng, Y.: Magnetic separation of polymer hybrid iron oxide nanoparticles triggered by temperature. Chemical Communications. 2765–2767 (2006). https://doi.org/10.1039/B604202C

8. Balasubramaniam, S., Pothayee, N., Lin, Y., House, M., Woodward, R.C., St. Pierre, T.G., Davis, R.M., Riffle, J.S.: Poly(N-isopropylacrylamide)-Coated Superparamagnetic Iron Oxide Nanoparticles: Relaxometric and Fluorescence Behavior Correlate to Temperature-Dependent

Aggregation. Chemistry of Materials. 23, 3348–3356 (2011). https://doi.org/10.1021/cm2009048

9. Kurzhals, S., Zirbs, R., Reimhult, E.: Synthesis and Magneto-Thermal Actuation of Iron Oxide Core–PNIPAM Shell Nanoparticles. ACS Applied Materials & Interfaces. 7, 19342–19352 (2015). https://doi.org/10.1021/acsami.5b05459

10. Reimhult, E., Schroffenegger, M., Lassenberger, A.: Design Principles for Thermoresponsive Core–Shell Nanoparticles: Controlling Thermal Transitions by Brush Morphology. Langmuir. 35, 7092–7104 (2019). https://doi.org/10.1021/acs.langmuir.9b00665

11. Zhang, Q., Weber, C., Schubert, U.S., Hoogenboom, R.: Thermoresponsive polymers with lower critical solution temperature: from fundamental aspects and measuring techniques to recommended turbidimetry conditions. Materials Horizons. 4, 109–116 (2017). https://doi.org/10.1039/C7MH00016B

12. Heskins, M., Guillet, J.E.: Solution Properties of Poly(N-isopropylacrylamide). Journal of Macromolecular Science: Part A - Chemistry. 2, 1441–1455 (1968). https://doi.org/10.1080/10601326808051910

13. Avraham, H., Martin, K., M., W.F.: Poly(N-isopropylacrylamide) Phase Diagrams: Fifty Years of Research. Angewandte Chemie International Edition. 54, 15342–15367 (2015). https://doi.org/10.1002/anie.201506663

14. Kano, M., Kokufuta, E.: On the Temperature-Responsive Polymers and Gels Based on N-Propylacrylamides and N-Propylmethacrylamides. Langmuir. 25, 8649–8655 (2009). https://doi.org/10.1021/la804286j

15. Hirano, T., Nakamura, K., Kamikubo, T., Ishii, S., Tani, K., Mori, T., Sato, T.: Hydrogen-bond-assisted syndiotactic-specific radical polymerizations of N-alkylacrylamides: The effect of the N-substituents on the stereospecificities and unusual large hysteresis in the phase-transition behavior of aqueous solution of syndiotactic poly(. Journal of Polymer Science Part A: Polymer Chemistry. 46, 4575–4583 (2008). https://doi.org/https://doi.org/10.1002/pola.22797

16. Graillot, A., Bouyer, D., Monge, S., Robin, J.-J., Loison, P., Faur, C.: Sorption properties of a new thermosensitive copolymeric sorbent bearing phosphonic acid moieties in multi-component solution of cationic species. Journal of Hazardous Materials. 260, 425–433 (2013). https://doi.org/https://doi.org/10.1016/j.jhazmat.2013.05.050

17. Queffélec, C., Petit, M., Janvier, P., Knight, D.A., Bujoli, B.: Surface Modification Using Phosphonic Acids and Esters. Chemical Reviews. 112, 3777–3807 (2012). https://doi.org/10.1021/cr2004212

18. Paniagua, S.A., Giordano, A.J., Smith, O.L., Barlow, S., Li, H., Armstrong, N.R., Pemberton, J.E., Brédas, J.-L., Ginger, D., Marder, S.R.: Phosphonic Acids for Interfacial Engineering of Transparent Conductive Oxides. Chemical Reviews. 116, 7117–7158 (2016). https://doi.org/10.1021/acs.chemrev.6b00061

19. Chanteau, B., Fresnais, J., Berret, J.-F.: Electrosteric Enhanced Stability of Functional Sub-10 nm Cerium and Iron Oxide Particles in Cell Culture Medium. Langmuir. 25, 9064–9070 (2009). https://doi.org/10.1021/la900833v

20. Torrisi, V., Graillot, A., Vitorazi, L., Crouzet, Q., Marletta, G., Loubat, C., Berret, J.-F.: Preventing Corona Effects: Multiphosphonic Acid Poly(ethylene glycol) Copolymers for Stable Stealth Iron Oxide Nanoparticles. Biomacromolecules. 15, 3171–3179 (2014). https://doi.org/10.1021/bm500832q

21. Ramniceanu, G., Doan, B.-T., Vezignol, C., Graillot, A., Loubat, C., Mignet, N., Berret, J.-F.: Delayed hepatic uptake of multi-phosphonic acid poly(ethylene glycol) coated iron oxide measured by real-time magnetic resonance imaging. RSC Advances. 6, 63788–63800 (2016). https://doi.org/10.1039/C6RA09896G

22. Baldim, V., Bia, N., Graillot, A., Loubat, C., Berret, J.-F.: Monophosphonic versus Multiphosphonic Acid Based PEGylated Polymers for Functionalization and Stabilization of Metal (Ce, Fe, Ti, Al) Oxide Nanoparticles in Biological Media. Advanced Materials Interfaces. 6, 1801814 (2019). https://doi.org/10.1002/admi.201801814

23. Grein-Iankovski, A., Graillot, A., Radiom, M., Loh, W., Berret, J.-F.: Template-Free Preparation of Thermoresponsive Magnetic Cilia Compatible with Biological Conditions. The Journal of Physical Chemistry C. 124, 26068–26075 (2020). https://doi.org/10.1021/acs.jpcc.0c09089

24. Graillot, A., Monge, S., Faur, C., Bouyer, D., Duquesnoy, C., Robin, J.-J.: How to easily adapt cloud points of statistical thermosensitive polyacrylamide-based copolymers knowing reactivity ratios. RSC Advances. 4, 19345–19355 (2014). https://doi.org/10.1039/C4RA00140K

25. Massart, R., Dubois, E., Cabuil, V., Hasmonay, E.: Preparation and properties of monodisperse magnetic fluids. Journal of Magnetism and Magnetic Materials. 149, 1–5 (1995). https://doi.org/10.1016/0304-8853(95)00316-9

26. Berret, J.-F., Sandre, O., Mauger, A.: Size Distribution of Superparamagnetic Particles Determined by Magnetic Sedimentation. Langmuir. 23, 2993–2999 (2007). https://doi.org/10.1021/la061958w

27. Giamblanco, N., Marletta, G., Graillot, A., Bia, N., Loubat, C., Berret, J.-F.: Serum Protein-Resistant Behavior of Multisite-Bound Poly(ethylene glycol) Chains on Iron Oxide Surfaces. ACS Omega. 2, 1309–1320 (2017). https://doi.org/10.1021/acsomega.7b00007

28. Schneider, C.A., Rasband, W.S., Eliceiri, K.W.: NIH image to ImageJ: 25 years of image analysis. Nature methods. 9, 671–675 (2012)

29. Chevry, L., Sampathkumar, N.K., Cebers, A., Berret, J.-F.: Magnetic wire-based sensors for the microrheology of complex fluids. Physical Review E. 88, 62306 (2013). https://doi.org/10.1103/PhysRevE.88.062306

30. Qi, L., Sehgal, A., Castaing, J.-C., Chapel, J.-P., Fresnais, J., Berret, J.-F., Cousin, F.: Redispersible Hybrid Nanopowders: Cerium Oxide Nanoparticle Complexes with Phosphonated-PEG Oligomers. ACS Nano. 2, 879–888 (2008). https://doi.org/10.1021/nn700374d

31. Graillot, A., Bouyer, D., Monge, S., Robin, J.-J., Faur, C.: Removal of nickel ions from aqueous solution by low energy-consuming sorption process involving thermosensitive copolymers with phosphonic acid groups. Journal of Hazardous Materials. 244–245, 507–515 (2013). https://doi.org/10.1016/J.JHAZMAT.2012.10.031

32. Zhao, C., Ma, Z., Zhu, X.X.: Rational design of thermoresponsive polymers in aqueous solutions: A thermodynamics map. Progress in Polymer Science. 90, 269–291 (2019). https://doi.org/https://doi.org/10.1016/j.progpolymsci.2019.01.001

33. Zhulina, E.B., Birshtein, T.M., Borisov, O. V: Curved polymer and polyelectrolyte brushes beyond the Daoud-Cotton model. The European Physical Journal E. 20, 243–256 (2006). https://doi.org/10.1140/epje/i2006-10013-5

34. Shan, J., Chen, J., Nuopponen, M., Tenhu, H.: Two Phase Transitions of Poly(N-

isopropylacrylamide) Brushes Bound to Gold Nanoparticles. Langmuir. 20, 4671–4676 (2004). https://doi.org/10.1021/la0363938

35. Schroffenegger, M., Reimhult, E.: Thermoresponsive Core-Shell Nanoparticles: Does Core Size Matter? Materials. 11, 1654 (2018). https://doi.org/10.3390/ma11091654

36. Zhao, J., Hoogenboom, R., Van Assche, G., Van Mele, B.: Demixing and Remixing Kinetics of Poly(2-isopropyl-2-oxazoline) (PIPOZ) Aqueous Solutions Studied by Modulated Temperature Differential Scanning Calorimetry. Macromolecules. 43, 6853–6860 (2010). https://doi.org/10.1021/ma1012368

37. Cheng, H., Shen, L., Wu, C.: LLS and FTIR Studies on the Hysteresis in Association and Dissociation of Poly(N-isopropylacrylamide) Chains in Water. Macromolecules. 39, 2325–2329 (2006). https://doi.org/10.1021/ma052561m

38. Carneiro, N.M., Percebom, A.M., Loh, W.: Quest for Thermoresponsive Block Copolymer Nanoparticles with Liquid–Crystalline Surfactant Cores. ACS Omega. 2, 5518–5528 (2017). https://doi.org/10.1021/acsomega.7b00905

39. Brinatti, C., Akhlaghi, S.P., Pires-Oliveira, R., Bernardinelli, O.D., Berry, R.M., Tam, K.C., Loh, W.: Controlled coagulation and redispersion of thermoresponsive poly di(ethylene oxide) methyl ether methacrylate grafted cellulose nanocrystals. Journal of Colloid and Interface Science. 538, 51–61 (2019). https://doi.org/https://doi.org/10.1016/j.jcis.2018.11.071

40. Katsumoto, Y., Tsuchiizu, A., Qiu, X., Winnik, F.M.: Dissecting the Mechanism of the Heat-Induced Phase Separation and Crystallization of Poly(2-isopropyl-2-oxazoline) in Water through Vibrational Spectroscopy and Molecular Orbital Calculations. Macromolecules. 45, 3531–3541 (2012). https://doi.org/10.1021/ma300252e

41. Wedel, B., Hertle, Y., Wrede, O., Bookhold, J., Hellweg, T.: Smart Homopolymer Microgels: Influence of the Monomer Structure on the Particle Properties, (2016)

42. de Oliveira, T.E., Marques, C.M., Netz, P.A.: Molecular dynamics study of the LCST transition in aqueous poly(N-n-propylacrylamide). Phys. Chem. Chem. Phys. 20, 10100–10107 (2018). https://doi.org/10.1039/C8CP00481A

43. Blackman, L.D., Gibson, M.I., O'Reilly, R.K.: Probing the causes of thermal hysteresis using tunable Nagg micelles with linear and brush-like thermoresponsive coronas. Polymer Chemistry. 8, 233–244 (2017). https://doi.org/10.1039/C6PY01191H

44. Yan, M., Fresnais, J., Berret, J.-F.: Growth mechanism of nanostructured superparamagnetic rods obtained by electrostatic co-assembly. Soft Matter. 6, 1997–2005 (2010). https://doi.org/10.1039/B919992F

45. Tierno, P.: Recent advances in anisotropic magnetic colloids: realization, assembly and applications. Physical Chemistry Chemical Physics. 16, 23515–23528 (2014). https://doi.org/10.1039/C4CP03099K